\documentclass[aps,prx,showpacs,a4paper,twocolumn]{revtex4-1}
\usepackage{graphicx}
\usepackage{color}
\usepackage{amsmath,amssymb,amsopn,comment}
\usepackage{bbm}
\usepackage{dsfont}
 
\def\bra#1{\mathinner{\langle{#1}|}}
\def\ket#1{\mathinner{|{#1}\rangle}}

\def\prjct#1{\mathinner{|{#1}\rangle}\!\!\mathinner{\langle{#1}|}}
\newcommand{\coh}[2]{\mathinner{|{#1}\rangle}\!\!\mathinner{\langle{#2}|}}
\newcommand{\braket}[2]{\langle #1  |#2\rangle}

\def\text#1{\textrm{#1}}
\def\id{\mathds{1}}

\def\eq{\begin{equation}}
\def\eeq{\end{equation}}
\def\tr{\text{tr}}

\def\P{\mathcal{P}}
\newcommand{\one}{\mbox{$1 \hspace{-1.0mm}  {\bf l}$}}

\begin{document}
\title{No-signaling bounds for quantum cloning and metrology}
\author{P. Sekatski, M. Skotiniotis, W. D\"ur}
\affiliation{Institut f\"ur Theoretische Physik, Universit\"at Innsbruck, Technikerstr. 25, A-6020 Innsbruck, Austria}
\date{\today}

\begin{abstract}
The impossibility of superluminal communication is a fundamental principle of physics. 
Here we show that this principle underpins the performance of several fundamental tasks in quantum information 
processing and quantum metrology. In particular, we derive tight no-signaling bounds for probabilistic cloning and 
super-replication that coincide with the corresponding optimal achievable fidelities and rates known. 
In the context of quantum metrology, we derive the Heisenberg limit from the no-signaling principle for 
certain scenarios including reference frame alignment and maximum likelihood state estimation.  
We elaborate on the equivalence of assymptotic phase-covariant cloning and phase estimation for different figures of merit. 
\end{abstract}
\maketitle
\section{Introduction}

Nothing can travel faster than the speed of light. This is one of the pillars of modern physics and an explicit element of 
Einstein's theory of relativity.  Any violation of this principle would lead to problems with local causality giving rise to 
logical contradictions. This principle not only applies to matter, but also to information, rendering superluminal 
communication impossible. Whilst not explicitly contained in the postulates of quantum mechanics all attempts to 
construct or observe violations of this principle have failed, leading us to believe that this is indeed a basic ingredient 
of our description of nature. In fact, modifications of quantum mechanics, e.g., by allowing non-linear dynamics, would 
lead to signaling and a violation of this fundamental principle~\cite{Gisin:89, *Gisin:90, Polchinski:91, Simon:01}. It is 
therefore natural to assume that no-signaling holds and try to deduce what follows under such an assumption.

Indeed, the no-signaling principle has been used to derive bounds and limitations on several physical processes and 
tasks. These include the observation that a perfect quantum copying machine would allow for superluminal 
communication~\cite{Herbert:82, Ghirardi:79, Simon:99}, limitations on universal quantum $1 \to 2$ 
cloning~\cite{Gisin:98, Ghosh:99} and $1 \to M$ cloning~\cite{Gedik:13}, a security proof for quantum 
communication~\cite{Barrett:05}, optimal state discrimination~\cite{Bae:11}, and bounds on 
the success probability of port-based teleportation~\cite{Garcia:13}. However, no-signaling alone is not 
restrictive enough as it allows for stronger non-local correlations than possible within quantum 
mechanics~\cite{Popescu:94}, and several attempts have been made to further supplement the no-signaling principle 
in order to retrieve quantum mechanical correlations~\cite{Brassard:06,*Linden:07,* 
Skrzypczyk:09,Pawlowski:09,Navascues:10, Fritz:13}.

Here we derive limitations on optimal quantum strategies from fundamental principles. In particular we show
\begin{itemize}
\item Tight no-signaling bound on probabilistic phase-covariant quantum cloning.
\item Asymptotically tight no-signaling bound on unitary super-replication. 
\item A derivation of the Heisenberg limit for metrology from the no-signaling condition.
\item Equivalence between asymptotic quantum cloning and phase estimation.
\item Quantum protocols that achieve the bounds placed by no-signaling.  
\end{itemize}
We assume the Hilbert space structure of pure states and show how the no-signaling principle directly leads to 
tight bounds on different fundamental tasks in quantum information processing and quantum metrology. 
We start by showing how the impossibility of faster-than-light communication between Alice and Bob can be used to 
provide upper bounds on Bob's ability to perform certain tasks, even if Bob has access to supra-quantum 
resources. Not only does the no-signalling principle allow us to prove ultimate limits on these 
fundamentally important tasks, it also allows us to demonstrate the optimality of known protocols and shed light on 
the recently discovered possibility of probabilistic super-replication of states~\cite{Ch:13} and 
operations~\cite{DSS:14, Ch:15}. 
  
We derive a no-signaling bound on the global fidelity of $N\to M$ probabilistic phase-covariant
cloning ~\cite{Ch:13}.  Our derivation is constructive and we provide the optimal deterministic quantum protocol that 
achieves the bound~\cite{Ch:13}.  In similar fashion, we derive a no-signaling bound on the replication of unitary 
operations~\cite{DSS:14}, which is tight in the large $M$ limit.
Furthermore, we derive the Heisenberg limit of quantum metrology solely from the no-signaling principle, more specifically for phase reference 
alignment~\cite{Buzek:99, Berry:00, Ch:04}. We find a tight 
no-signaling bound on the maximal likelihood and a bound with the correct scaling on the fidelity of reference 
frame alignment for phase both for the uniform prior as well as for a non-uniform prior probability distribution. 

We show that the no-signaling condition can be used to establish bounds on the 
performance of quantum information tasks for which no bounds are known, or for which the brute force optimization of 
the tasks is hard. This demonstrates an alternative approach to establish the possibilities and limitations of quantum information 
processing, which is based on fundamental principles rather than actual protocols. We emphasize that this approach is not limited 
to the specific tasks discussed here, but is generally applicable. 

We also discuss the correspondence between asymptotic phase-covariant quantum cloning and state estimation for different 
figures of merit, solving the open problem of whether asymptotic cloning, quantified by the global fidelity, is equivalent to state 
estimation~\cite{Yang:13}. Finally we supplement our approach by a general argument, extending that of~\cite{Simon:01}, showing 
that optimal quantum protocols are at the edge of no-signaling.

\section{No-signaling}
\label{sec:no-signaling}

In this section we describe the operational setting underpinning all three tasks we consider (cloning, replication of 
unitaries, and metrology), as well as the no-signalling condition. All three tasks we consider can 
be described in the following operationally generic setting. A party, Bob, possess an $N$-qubit state, 
\eq
\ket{\Phi^N}_B = \sum_{\bf v} a_{\bf v} \ket{\bf v}_B =\sum_{n=0}^N p_n \underbrace{\sum_{|{\bf v}|=n}\frac{a_{\bf v}}
{p_n} \ket{\bf v}_B}_{\ket{\tilde n}_B},
\eeq
where ${\bf v}$ runs over all $N$-bit stringsand $\ket{\tilde n}_B$ is a superposition over all states with Hamming 
weight $|{\bf v}|=n$..  Bob then receives, via a remote preparation scenario to be described shortly, the 
action of a 
unitary operator $U_\theta^{\otimes N}$ such that 
\begin{equation}
\ket{\Phi^N_\theta}=U_\theta^{\otimes N}\ket{\Phi^N}, 
\label{state}
\end{equation}
where $U_\theta=e^{i\theta H}$ with $H$ an arbitrary Hamiltonian acting on $2$-level systems (qubits) with 
\emph{spectral radius} $\sigma(H)$, $\theta$ uniformly chosen from $(0,2\pi\sigma(H)]$.  

Bob has to process $U_\theta^{\otimes N}$ for some quantum information task in an optimal way.   In particular, we do not 
demand that Bob's processing be described by linear maps, nor do we demand that the mapping from valid quantum states to 
probability distributions be given by the Born rule.  All that we require of Bob's processing outcomes is that they should be valid 
inputs for someone whose processing power is limited by quantum theory. We choose such a setting because our goal is rather 
pragmatic---we wish to derive upper bounds on \emph{quantum} information tasks.  Hence, throughout this work we shall 
assume that all of Bob's \emph{static} resources, i.e., pure states of physical systems, ensembles of pure states, and probability 
distributions, are described 
within the framework of quantum theory, but Bob's \emph{dynamical} resources, i.e., processing maps, are not.
In fact imposing no-signaling condition for quantum static resources is equivalent 
to imposing quantum mechanics (see Sec.\ref{sec: Discussion}), but when a direct optimization over quantum strategies is 
unfeasible the no-signaling argument can help to show that a known strategy is optimal. 

What Bob has to output varies depending on which 
task he performs.  For example, if the required task is the cloning of the state $\ket{\psi(\theta)}$, then 
Bob has to output an $M$-qubit state, $\rho^M_\theta$, that is a close approximation to $\ket{\psi(\theta)}^{\otimes M}$. If 
the required task is the replication of the unitary operator $U_\theta$ then Bob has to output a quantum channel 
acting on the Hilbert space of $M$ qubits that is a close approximation to $U_\theta^{\otimes M}$.  Finally, if the 
required task is to estimate the parameter $\theta$, then Bob must output a probability distribution corresponding to  
his updated knowledge about parameter $\theta$.  We denote the outcome of Bob's processing, be it a quantum 
state, channel, or probability distribution, by $\P(\theta)$. 

\begin{figure}[ht!]
\includegraphics[width=8cm]{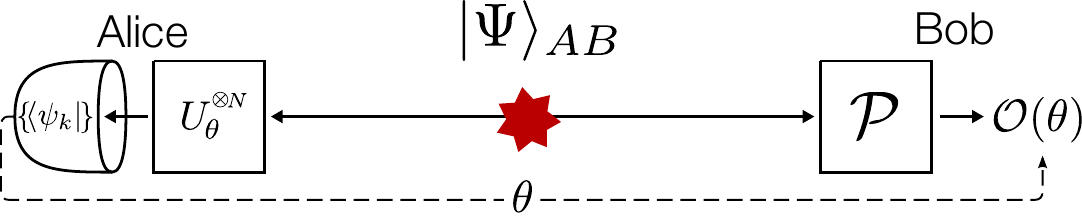}
\caption{Generic setting for faster-than-light communication. Alice and Bob share an entangled state $\ket{\Psi}_{AB}$. By applying $U_\theta^{\otimes N}$ followed by a suitable measurement Alice can prepare any ensemble $\{ p_k \prjct{\Phi_{\theta|k}} \}$, which Bob processes into $\mathcal{O}(\theta)=\sum_k p_k \mathcal{P}(\theta|k)$. The no-signaling condition imposes that $\mathcal{O(\theta)}$ is independent of $\theta$ chosen by Alice.}
\label{fig1}
\end{figure}

To incorporate the no-signaling condition we consider that Bob holds one part of a suitably chosen entangled state
\eq
\ket{\Psi}_{AB} = \sum_{n=0}^N c_n \ket{n}_A \ket{\tilde n}_B
\eeq
which he shares with Alice, where Alice keeps the $(N+1)$-level system spanned by $\{\ket{n}_A\}$ and Bob holds 
the $2^N$-dimensional system spanned by $\{\ket{\tilde n}_B\}$. The state $\ket{\Psi}_{AB}$ can always be chosen 
such that Bob receives the action of $U^{\otimes N}_\theta$ on an arbitrary input state.  This is achieved by Alice 
first performing $(U^{\otimes N}_\theta\otimes \one) \ket{\Psi}_{AB}$, followed by a 
measurement in the \emph{Fourier basis} $\{ \ket{ k} \propto \sum_n e^{i n \frac{2\pi k }{N+1}} \ket{n}_A\}$ 
with $k = 0, \ldots, N$ (see Fig.~\ref{fig1}).  If Alice obtains outcome $k$ then Bob's state becomes
\begin{equation}
\ket{\Phi^N_{\theta+\frac{2\pi\,k}{N+1}}}= U_{\theta+\frac{2\pi\, k}{N+1}}^{\otimes N}\ket{\Phi^N}. 
\end{equation}
As all outcomes, $\ket{k}$, are equally likely Bob ends up with a random state from the ensemble 
$\{\ket{\Phi^N_{\theta+\frac{2\pi\,k}{N+1}}}, \, k\in(0,\ldots, N)\}$. 

The no-signaling condition now requires that Bob, who does not know which unitary $U_\theta, \, \theta\in(0,2\pi]$ 
was chosen by Alice, can not learn $\theta$ from the above ensemble no matter what processing power, quantum or 
otherwise, Bob has at his disposal. If this were not the case then Alice and Bob, 
who are spatially separated, can use the above construction to perform faster-than-light communication. 
Denoting the outcome of Bob's processing by $\mathcal{P}(\theta | k)$ the no-signaling condition requires that the 
mixture
\eq\label{ns general}
\mathcal{O}(\theta)  =\frac{1}{N+1}\sum_{k=0}^{N} \mathcal{P}(\theta | k)
\eeq  
is independent of $\theta$ chosen by Alice.

Note that the no-signaling bound derived above is based on a particular way to embed a quantum information processing 
task into a communication scenario. The bound turns out to be tight in the present context but is not in general. We will come 
back to this point from a more general perspective in the Sec.~\ref{sec: Discussion}.

\section{Probabilistic phase covariant cloning}
\label{sec:PCC}

We first apply the no-signaling condition to the case of phase-covariant quantum cloning (PCC).  The latter task 
involves cloning an unknown state from the set $\{\ket{\psi(\theta)}=U(\theta)\ket\psi)\}$~\cite{Bruss:00, *D'Ariano:03,Ch:13}.  
We focus on PCC of equatorial states, $\ket{\psi(\theta)}=1/\sqrt{2}(\ket{0}+e^{i\theta}\ket{1})$,
which play a crucial role in proving the security of quantum key distribution~\cite{Barrett:05}.  Specifically, we provide a 
bound for the optimal PCC of $N$ qubits into $M>N$ qubits and show that this bound 
is achievable by a deterministic quantum mechanical strategy, if one drops the restriction of separable $N$-qubit input states.
The latter strategy involves the use of a suitable  $N$-partite entangled input state on which $U_\theta^{\otimes N}$ is 
applied.  We then show how our deterministic 
strategy is equivalent to the probabilistic PCC of~\cite{Ch:13}, by introducing a suitable filter 
operation that maps $\ket{\psi(\theta)}^{\otimes N}$ to the suitable $N$-partite entangled state.  

A \emph{deterministic}, phase-covariant quantum cloning machine is some transformation, $\mathcal{C}$, 
whose input is $N$ copies of an unknown equatorial qubit state $\ket{\psi(\theta)}$, that outputs an $M$-qubit state  
$\rho^M(\theta) =\mathcal{C}\left((\prjct{\psi(\theta)})^{\otimes N}\right)$.  Optimal deterministic cloning machines, be it 
state-dependent~\cite{Bruss:00, *D'Ariano:03, Wo:82} or state-independent~\cite{Buzek:96, Gisin:97, Bruss:98, Werner:98},  
have been constructed and tight bounds, for the case of $1\to 2$ cloning, based on the no-signaling condition have 
been derived~\cite{Gisin:98, Ghosh:99}. A probabilistic cloning machine is more powerful in 
that it allows for a much higher number of copies at the cost of succeeding only some of the time.  Indeed, probabilistic 
PCC, when successful, can output up to $N^2$ faithful copies of $\ket{\psi(\theta)}$.
However, the probability of success is exponentially small~\cite{Ch:13}.

If the input state to the probabilistic PCC machine is remotely prepared by Alice, as explained in Sec.~\ref{sec:no-signaling}, 
then the no-signaling condition on the output of Bob's probabilistic PCC procedure has to be 
independent of $\theta$, i.e.,  
\eq\label{ns_for_cloning}
\rho^M = \rho^M(\theta)  =\frac{1}{N+1}\sum_{k\geq 0}^N \rho_{\theta+ \frac{2\pi \,k}{N+1}}^M.
\eeq  
Following~\cite{Ch:13}, we quantify the success of the cloning procedure by the \emph{worst case global cloning fidelity} 
\eq\label{mf}
F_{wc}^{\mathcal C_{N\to M}} = \inf_\theta  F_C(\rho^M_\theta, (\prjct{\psi(\theta)})^{\otimes M}),
\eeq
where $F_C(\rho_\theta^M, (\prjct{\psi(\theta)})^{\otimes M})=\mathrm{Tr}\left(\rho_\theta^M (\prjct{\psi(\theta)})^{\otimes M}\right)$
is the \emph{global} fidelity between the output of the cloner and $M$ perfect copies of the input state.  

Recalling that the Uhlmann fidelity, $F_U(\rho,\sigma)= \tr \sqrt{\sqrt{\sigma} \rho \sqrt{\sigma}}$ it follows that  
$F_C(\rho_\theta^M, (\prjct{\psi(\theta)})^{\otimes M}) =F_U(\rho_\theta^M, (\prjct{\psi(\theta)})^{\otimes M})^2$.  
Moreover, as the worst case fidelity is always smaller or equal than the mean fidelity the following bound holds
\begin{align}\nonumber
F_{wc}^{\mathcal C_{N\to M}}\leq \left( \int \frac{d\theta}{2\pi}  F_U(\rho^M_\theta, (\prjct{\psi(\theta)})^{\otimes M}) \right)^2. 
\end{align}
Thus an upper bound for the worst case global cloning fidelity can be obtained by obtaining an upper bound on the mean 
Uhlmann fidelity. 

In order to upper bound the mean Uhlmann fidelity we first rewrite the latter as
\begin{align}\label{Fidelity2} 
&\int \frac{d\theta}{2\pi}  F_U(\rho^M_\theta, (\prjct{\psi(\theta)})^{\otimes M})=
\int_0^{2\pi}\sum_{k=0}^N \frac{d\theta}{2\pi(N+1)} \nonumber \times\\ \nonumber
& F_U(\rho^M_{\theta+\frac{2\pi k}{N+1}},
(U_{\frac{2\pi k}{N+1}}\prjct{\psi(\theta)}U^{\dagger}_{\frac{2\pi k}{N+1}})^{\otimes M})\\
&\leq \int_0^{2\pi} \frac{d\theta}{2\pi} F_U\left(\rho^M, \sum_{k=0}^N\frac{(U_{\frac{2\pi k}{N+1}}\prjct{\psi(\theta)}
U^{\dagger}_{\frac{2\pi k}{N+1}})^{\otimes M}}{N+1}\right),
\end{align}
where we have used the joint concavity of the Uhlmann fidelity, $F_U\left(\sum_i p_i\,\rho_i, \sum_i p_i\sigma_i\right)\geq
\sum_i p_i F_U(\rho_i,\sigma_i)$ in the last line of Eq.~\eqref{Fidelity2}. As 
$\ket{\psi(\theta)}=U_\theta\ket{+}$, and $[U_{\frac{2\pi k}{N+1}},U_\theta]=0$,  unitary invariance of the fidelity, 
$F(\rho ,U\sigma U^\dagger)=F(U^\dag \rho U,\sigma)$, allows us to shift the action of $U(\theta)^{\otimes M}$ onto 
$\rho^M$ and the integrand of Eq.~\eqref{Fidelity2} reads 
\eq
\int_0^{2\pi} \frac{d\theta}{2\pi} F_U\left(U_\theta^\dag \rho^M U_\theta, \sum_{k=0}^N\frac{(U_{\frac{2\pi k}{N+1}}\prjct{+}
U^{\dagger}_{\frac{2\pi k}{N+1}})^{\otimes M}}{N+1}\right).
\eeq
Finally using the concavity of the Uhlmann fidelity, 
$F\left(\sum_i p_i\,\rho_i, \sigma\right)\geq \sum_i p_i F(\rho_i, \sigma)$), to move the integral over $\theta$ inside the 
argument for the Uhlmann fidelity,  and defining the maps 
$\mathcal{G}_{\mathbb{Z}_{N+1}}[\cdot]\equiv\frac{1}{N+1}\sum_{k=0}^NU^{\otimes M}_{\frac{2\pi k}{N+1}}\,
(\cdot)\, U^{\dagger\,\otimes M}_{\frac{2\pi k}{N+1}}$ and $\mathcal{G}^{\dagger}_{\mathbb{U}(1)}[\cdot]\equiv \frac{1}{2\pi}
\int_0^{2\pi}\mathrm{d}\theta\, U^{\dagger\,\otimes M}_\theta \,(\cdot)\, U^{\otimes M}_\theta$, we obtain the desired upper 
bound for the worst case global cloning fidelity
\eq\label{Fidelity4}
F_{wc}^{\mathcal C_{N\to M}} \leq  F_U\left(\mathcal{G}^\dagger_{\mathbb{U}(1)}[\rho^M],\mathcal{G}_{\mathbb{Z}_{N+1}}
[(\prjct{+})^{\otimes M}]\right)^2.
\eeq

We now proceed to give an explicit expression for the upper bound of Eq.~\eqref{Fidelity4}.  
The maps $\mathcal{G}$ impose a 
\emph{block-diagonal} structure on any density matrix on which they act, making it it easy to find $\rho_M$  that maximizes 
Eq.~\eqref{Fidelity4}. As the state $(\prjct{+})^{\otimes M}$ is symmetric under permutations it suffices to maximize over all 
symmetric $\rho^M$. For any permutation symmetric $\rho^M$, $\mathcal{G}^{\dagger}_{\mathbb{U}(1)}[\rho^M]$ is diagonal 
and can be written as
\eq\label{block1}
\mathcal{G}^\dagger_{\mathbb{U}(1)}[\rho^M] = \bigoplus_{n=0}^M p_n \prjct{n,M},
\eeq
where $\{\ket{n,M}\}_{n=0}^{M}$ is an orthonormal basis  spanning the symmetric subspace of $M$ qubits with $n$ qubits in 
state
$\ket{1}$ and $M-n$ qubits in state $\ket{0}$.  Correspondingly, we may write 
\eq\label{block2}
\mathcal{G}_{\mathbb{Z}_{N+1}}[(\prjct{+})^{\otimes M}]=\bigoplus_{\lambda=0}^N\prjct{\phi^{(\lambda)}},
\eeq
where $\ket{\phi^{(\lambda)}}=\sum_{n |\, n\mod (N+1) =\lambda}\sqrt{\frac{\binom{M}{n}}{2^M}}\ket{n,M}$  are unnormalized 
pure symmetric states with the sum running over all $n$ that have a reminder $\lambda$ after division by $N+1$.
The states $\ket{\phi^{(\lambda)}}$ have non-zero overlap with the symmetric states $\ket{n,M}$ whenever 
$n \mod (N+1) =\lambda$.

Because of the block-diagonal structure we rewrite the mean Uhlmann fidelity as
\begin{align}
F_U(\mathcal{G}^\dagger_{\mathbb{U}(1)}[\rho^M],\mathcal{G}_{\mathbb{Z}_{N+1}}[(\prjct{+})^{\otimes M}])= \nonumber\\
\sum_{\lambda=0}^N \sqrt{\bra{\phi^{(\lambda)}} \mathcal{G}_{\mathbb{U}(1)}[\rho^M] \ket{\phi^{(\lambda)}}}.
\end{align}
Denoting by $p_\lambda = \sum_{\{n| n \mod (N+1) =\lambda\}} p_n$ the probability of projecting $\mathcal{G}^\dagger_{\mathbb{U}(1)}[\rho^M]$ on the sector with a given  $\lambda$ we can maximize the Uhlmann fidelity by  
optimizing each sector $\lambda$ independently.  This is achieved by finding the $n$ in each sector $\lambda$ such that the overlap $\braket{\phi^{(\lambda)}}{n,M}$ is maximized. The maximum Uhlmann fidelity then reads 
\begin{align}
\max_{p_n} F_U(\bigoplus_{n=0}^M p_n \prjct{n,M},\bigoplus_{\lambda=0}^N\prjct{\phi^{(\lambda)}} ) =\nonumber\\
\max_{p_\lambda} \sqrt{p_\lambda \max_n |\braket{\phi^{(\lambda)}}{n,M}|^2}= \sqrt{\sum_\lambda \max_n  |
\braket{\phi^{(\lambda)}}{n,M}|^2}.
\end{align}

The probability $|\braket{\phi^{(\lambda)}}{n,M}|^2 = \frac{1}{2^M}\binom{M}{n}$  is given by the binomial distribution if 
$n\mod (N+1) = \lambda$ and is zero otherwise. Thus, it is always optimal to choose $n\mod (N+1)$ closest to $\frac{M}{2}$.  
Doing so for all $\lambda$ we find that the maximal fidelity is given by the square root of the sum of the $N+1$ largest terms of the 
binomial distribution $\frac{1}{2^M}\binom{M}{n}$. Hence, the upper bound for  the worst case global cloning fidelity reads
\eq\label{fidelity6}
F_{wc}^{\mathcal C_{N\to M}} \leq  \frac{1}{2^M}\sum_{\lambda=0}^N  \binom{M}{\lfloor \frac{M-N}{2}\rfloor + \lambda},
\eeq
where $\lfloor\cdot\rfloor$ denotes the floor function. Finally, noting that the binomial distribution, $\frac{1}{2^M}\binom{M}{n}$ can be approximated by a Gaussian $\mathcal{N}(\mu=M/2, \sigma=\sqrt{M}/2)$, the upper bound in Eq.~\eqref{fidelity6} can be approximated, for $M\gg N$, by 
 \begin{equation}
F_{wc}^{\mathcal C_{N\to M}}\leq \mathrm{erf}\left(\frac{N+1}{\sqrt{2 M}}\right). 
\label{asymptoticfidelity}
\end{equation}

We note that as long as $M=\mathcal{O}(N^2)$ the cloning fidelity approaches unity in the limit $N\to\infty$.  Indeed, 
one can make an even stronger claim.  Any replication procedure that respects the no-signaling condition and produces 
a number of replicas $M=\mathcal{O}(N^{2+\alpha})$ does so with a cloning fidelity that tends to zero exponentially 
fast.

We now show how one can achieve the no-signaling bound of Eq.~\eqref{asymptoticfidelity} using a deterministic 
quantum mechanical strategy. Instead of $N$-copies of $\ket{\psi(\theta)}$, suppose Bob prepares the entangled state 
\eq\label{deterministic}
\ket{\Phi^N} \propto \sum_{\lambda=0}^N \sqrt{ \binom{M}{\lfloor \frac{M-N}{2}\rfloor + \lambda}}\ket{N,\lambda}.
\eeq
Bob now applies the cloning map $\mathcal{C}: \ket{N,\lambda} \mapsto \ket{M,\lfloor \frac{M-N}{2}\rfloor + \lambda}$, 
that maps totally symmetric $N$-qubit states to totally symmetric $M$-qubit states.  This strategy achieves the bound 
of Eq.~\eqref{asymptoticfidelity} as the latter is valid for all input states. We pause to note that the above result does not 
contradict the well know limits for deterministic cloning, as in the latter  Bob is forced to input $N$-copies of a qubit state.

The bound of Eq.~\eqref{asymptoticfidelity} is the ultimate bound that can be achieved even by a probabilistic strategy.  
Indeed, the best probabilistic quantum mechanical PCC attains precisely the no-signaling bound of 
Eq.~\eqref{asymptoticfidelity}~\cite{Ch:13}. In fact there is an easy way to see how the probabilistic strategy 
of~\cite{Ch:13} and the deterministic strategy described above are related.  Starting from $N$ copies of the state 
$\ket{\psi(\theta)}$, the probabilistic PCC of~\cite{Ch:13} has Bob first apply the probabilistic filter that 
projects onto the state $\ket{\Phi^N}$ of Eq.~\eqref{deterministic} and succeeds with probability 
$p_{yes} = |\braket{\Phi^N}{+}^{\otimes N}|^2$. As such a filter commutes with the unitary $U_\theta^{\otimes N}$ it 
can be seen as part of the overall state preparation. 
The advantage, then, of probabilistic PCC can be simply 
understood as a passage from the standard quantum limit in quantum metrology, achieved for separable input states, 
to the Heisenberg limit achieved by entangled input states.  Notice that no probabilistic advantage exists for the case of $1\to M$ cloning.  For the latter, the fidelity of Eq.~\eqref{asymptoticfidelity} takes the simple form $F^{\mathcal C_{1\to M}} 
= \frac{1}{2^{M-1}}\binom{M}{\frac{M-1}{2}}$ for $M$ odd and $F^{\mathcal C_{1\to M}} = \frac{1}{2^{M}}\binom{M+1}
{\frac{M}{2}+1}$ for $M$ even and is known to be achievable by a deterministic strategy~\cite{D'Ariano:03}. 
\begin{figure}[ht!]
\includegraphics[width=6cm]{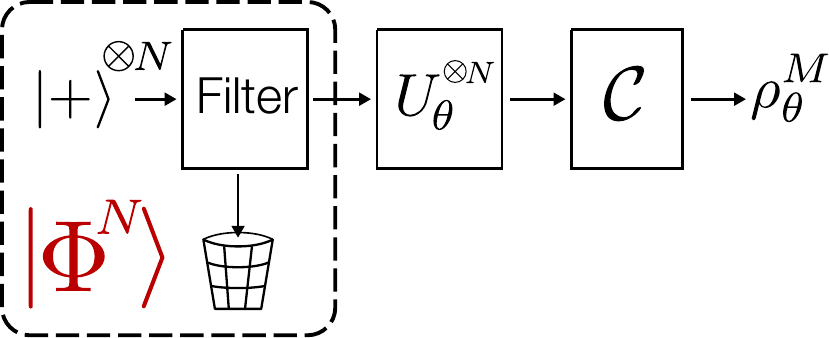}
\caption{Equivalence between probabilistic PCC and deterministic PCC using arbitrary states. The filter in the probabilistic cloning can be viewed as part of a probabilistic preparation of a general state from a separable N-qubit state. Allowing for arbitrary input states makes the preparation process deterministic.}
\label{fig2}
\end{figure}

\section{Replication of unitaries}

We now consider the task where Bob has to output an approximation, $V_\theta$, of $U^{\otimes M}_\theta$ having 
received only $N$ uses of the black box implementing the unitary transformation $U_\theta$~\cite{Ch:08, 
DSS:14}.  The figure of merit that one uses is the global Jamio\l{}kowski fidelity 
(process fidelity)~\cite{Gilchrist:05,*Dur:05},
\eq\label{processfidelity}
F(U^{\otimes M}_\theta, V_\theta) = \bra{\psi_{U^{\otimes M}_\theta}} \rho_{V_\theta} \ket{\psi_{U^{\otimes M}_\theta}},
\eeq
averaged over all $\theta$, where $\ket{\psi_{U^{\otimes M}_\theta}} = (\one\otimes U_\theta^{\otimes M}) \ket{\Phi^+}$ 
and 
$\rho_{V_\theta} = \one \otimes V_\theta (\prjct{\Phi^+})$, with $\ket{\Phi^+}=1/\sqrt{2^M}\sum_{n}\ket{n}\ket{n}$, where $n$ are the $M$-qubit bit strings, are the corresponding Choi-Jamio\l{}kowski states~\cite{Jamiolkowski:72, *Choi:72} for 
$U^{\otimes M}_\theta$ and $V_\theta$ respectively.  It was shown in~\cite{DSS:14} that when $M<N^2$ Bob can 
approximate $U_\theta^{\otimes M}$ almost perfectly, i.e., with process fidelity approaching unity in the large $N$ limit. 
We now show that the protocol in~\cite{DSS:14} saturates the no-signaling bound. 

In order to apply the no-signaling condition for the case of unitary replication in an easy way we consider the 
following communication scenario.    Alice prepares the Choi-Jamio\l{}kowski state corresponding to 
$U^{\otimes N}_\theta$, $\ket{\psi_{U^{\otimes N}_\theta}}$, at Bob's side, which he can then use to probabilistically implement $U^{\otimes N}_\theta$ on an arbitrary input state~\cite{Dur:01}. Consequently, the protocol for which we shall derive a no-signaling bound is inherently probabilistic.  We note that a 
bound for a probabilistic protocol is automatically a bound for a deterministic protocol as well, as the former are less 
restrictive than the latter.
 
The no-signaling constraint for unitary replication takes the form
\eq
\mathcal{R}_N^M = \frac{1}{N+1}\sum_{k=0}^{N} \id\otimes V_{\theta + \frac{2\pi k}{N+1}} (\prjct{\Phi^+}),
\eeq
and is independent of $\theta$. As the worst case process fidelity (Eq.~\eqref{processfidelity}) is identical to the worst 
case global cloning fidelity used for PCC (Eq.~\eqref{mf}) the no-signaling bound for probabilistic replication of unitaries reads
\eq\label{unitaryrepbound}
F_{wc}(U^{\otimes M}_\theta, V_\theta)  \leq \frac{1}{2^M} \sum_{\lambda=0}^N
 \binom{M}{\lfloor \frac{M-N}{2}\rfloor + \lambda}.
\eeq
This bound is achieved, in the limit of large $M$, by the deterministic strategy in~\cite{DSS:14}, for which the fidelity is independent of $\theta$. This implies that probabilistic processes offer no advantage in this case. Thus, the optimal deterministic replication of 
unitary operations allowed by quantum mechanics is at the edge of no-signaling. 

\section{Quantum Metrology}

We now apply the no-signaling condition to provide bounds for quantum metrology.  The latter task 
involves the use of $N$ systems, known as the probes, prepared in a suitable state $\ket{\psi}\in\mathcal{H}^{\otimes 
N}$, and subjected to a dynamical evolution described by a completely positive map, $\mathcal{E}_\theta$, 
that imprints the value of $\theta$ onto their state, i.e., $\rho_\theta=\mathcal{E}_\theta(\prjct{\psi})$.  Information about 
the value of $\theta$ is retrieved by a suitable measurement of the $N$ probes.  The goal in quantum metrology is to 
choose the initial state $\ket{\psi}$ and final measurement such that the value of $\theta$ can be inferred as precisely 
as possible.

If the $N$ quantum probes are prepared in a separable quantum state, i.e., $\ket{\psi}=\ket{\phi}^{\otimes N}$, then 
the mean square error with which $\theta$ can be estimated, optimizing over all allowable measurements, 
scales inversely proportional with $N$~\cite{GLM04}.  This limit is known as the \emph{standard quantum 
limit}.  If, however, the $N$ probes are prepared in a suitably entangled state, then the mean square error with 
which $\theta$ can be estimated scales inversely proportional with $N^2$~\cite{GLM04}.  This limit is known as the 
\emph{Heisenberg limit}.  By allowing for a probabilistic strategy, the Heisenberg limit in precision can be obtained 
even with separable states~\cite{Fiurasek:06, Gendra:13a, *Gendra:13b}.  Recently, it was shown that both the standard and 
Heisenberg limits are related with the maximum replication rates corresponding to a deterministic and probabilistic 
PCC strategies respectively~\cite{Ch:13, Gendra:14}.   

We now show how the no-signaling condition implies that the ultimate bound in precision 
for metrology is the Heisenberg limit, even if supra-quantum processing is allowed.  We shall consider two particular 
examples of Bayesian quantum metrology; phase alignment, where the relevant parameter to be 
estimated is the phase of a local oscillator, $\theta\in (0,2\pi]$, which is initially completely unknown~\cite{BRS:07} 
(Sec.~\ref{sec:metrology}), and phase diffusion where our prior knowledge of the parameter, initially described by a delta 
function around some value $\theta_0$, diffuses over time~\cite{Demkowicz:11} (Sec.~\ref{sec:diffusion}).
We stress that whereas analytical bounds for phase alignment are known, for phase diffusion bounds are known only for 
small number of probes~\cite{Demkowicz:11}. This is due to the fact that the optimal strategy is difficult to compute, even 
numerically.  Nevertheless, our no-signaling constraint allows us to place an upper bound on the optimal fidelity of estimation 
for asymptotically many probe systems.  We emphasize that a similar strategy can be applied to a variety of quantum 
information processing tasks, where limitations of the processes can be gauged by fundamental principles.

In Sec.~\ref{Assymptotic Cloning} we establish the relationship between optimal quantum cloning protocols and measure and 
prepare strategies.  In particular, we show that a measure and prepare strategy that maximizes the alignment fidelity is 
asymptotically equivalent to a quantum cloning machine that maximizes the {\it per copy} fidelity, whereas a measure and 
prepare strategy that optimizes the maximum likelihood of estimation is asymptotically equivalent to a quantum cloning 
machine that maximizes the global fidelity.  This latter result answers the open question of Yang {\it et al.}~\cite{Yang:13} as 
to whether asymptotic cloning, where the quality of the cloned state is quantified by the global fidelity, is equivalent to state 
estimation.

\subsection{Metrology with uniform prior}
\label{sec:metrology}
Consider the problem of phase alignment, i.e., estimating a completely unknown phase, $\theta$.  We will utilize 
two different ways of quantifying the precision of estimation of $\theta$:  the \emph{maximum likelihood} of a correct guess, 
$\mu=p(\theta|\theta)$~\cite{Ch:04}, and the \emph{fidelity} of alignment, given by the payoff function 
$f=\cos^2\left(\frac{\theta-\theta'}{2}\right)$~\cite{Berry:00}.  For the case of phase alignment the no-signaling condition 
(Eq.~\eqref{ns general}) takes the form 
\begin{align}\label{ns alignment}
p(\theta'|\theta)=\frac{1}{N+1}\sum_{k=0}^N p\left(\theta'| \theta + \frac{2\pi k}{N+1}\right)
\end{align}
and is independent of $\theta$ (the same holds for a measurement with discrete outcomes). Note that we make no 
assumptions on how Bob obtains the probability distribution of Eq.~\eqref{ns alignment}.  In 
particular we do not restrict Bob's processing to be quantum mechanical. We only require that the inputs and 
outputs to Bob's processing apparatus be valid quantum states and probability distributions respectively.  

\subsubsection{Maximal likelihood}
\label{sec: maximal likelihood}
For the case where the precision is quantified by the maximum likelihood the no-signaling bound (Eq.~\eqref{ns 
alignment}) gives $p(\theta|\theta) \leq  (N+1)p(\theta)$.  If the estimate $\theta'$ is unbiased, all 
outcomes are equally likely and the no-signaling bound takes the simple form $p(\theta|\theta) \leq N+1$. The bound is 
known to be achievable using the state~\cite{Ch:04}
\eq\label{ml state}
\ket{\Phi^N_{m.l.}}=\frac{1}{\sqrt{N+1}}\sum_n \ket{n}.
\eeq

\subsubsection{Alignement fidelity}
\label{sec: alignment fidelity}
For the case where the precision is quantified by the fidelity of alignment, for each choice of $\theta,\, \theta'$ the 
fidelity must be properly weighted by the \emph{joint probability distribution}, 
$p(\theta',\theta)=p(\theta'|\theta)p(\theta)$.  The average fidelity of alignment is thus 
\eq
\label{meanfid}
\bar{f}= \int \frac{d\theta}{2\pi} \int\mathrm{d}\theta'\cos^2\left(\frac{\theta-\theta'}{2}\right) p(\theta'|\theta)
\eeq
The probability distribution that both maximizes the average fidelity and is compatible with no-signaling is 
\begin{align}\label{step probability}
p(\theta'|\theta)=
\left\{\begin{array}{cl}
\frac{N+1}{2\pi} & \text{if}\quad |\theta' -\theta| \leq \frac{\pi}{N+1}\\
0 & \text{otherwise}
\end{array}\right.
\end{align}
as we now show.  

Our aim is to distribute the probability distribution of Eq.~\eqref{ns alignment} amongst $N+1$ terms subject to the 
constraint that $\int\mathrm{d}\theta' p(\theta')=1$ such that the average fidelity of Eq.~\eqref{meanfid} is maximized.  
Without loss of generality assume that $\theta\in \left(0,\frac{2\pi}{N+1}\right)$.  If this is not the case we can always 
relabel the measurement outcomes $k\in(0,\ldots,N)$ such that $\theta$ lies in $\left(0,\frac{2\pi}{N+1}\right)$.  As 
$\cos^2\left(\frac{\theta'-\theta}{2}\right)$ is largest when $\theta-\theta'=0$ the average fidelity is optimized by setting 
$p(\theta'|\theta+\frac{2\pi k}{N+1})=0$ for $k\neq 0$.  As this is true for all randomly chosen~$\theta$, and using the 
constraint $\int\mathrm{d}\theta'\, p(\theta')=1$, it follows that $p(\theta'|\theta)=\frac{N+1}{2\pi}$ for $|\theta'-\theta|\leq
\frac{\pi}{N+1}$ and zero everywhere else.

We now derive the maximum average fidelity (Eq.~\eqref{meanfid}) compatible with no-signaling.  As the conditional 
probability distribution $p(\theta'|\theta)$  of Eq.~\eqref{step probability} depends only on the difference 
$\theta'-\theta$ we may write the average fidelity as 
\eq
\bar f=1-\int_{-\pi}^{\pi}\frac{\mathrm{d}\theta}{2\pi}\int_{-\pi}^{\pi}\mathrm{d}\theta' p(\theta'-\theta)\sin^2\left(\frac{\theta-\theta'}{2}\right)
\label{fid1}
\eeq
where we have used the identity $\cos^2(x)=1-\sin^2(x)$.  As the integrand in Eq.~\eqref{fid1} depends only on the difference $\theta'-\theta$ we may define $\phi=\theta'-\theta$ and $\mathrm{d}\phi=\mathrm{d}\theta'$ so that 
\eq
\bar f=1-\int_{-\pi}^{\pi}\mathrm{d}\phi\, p(\phi)\sin^2\left(\frac{\phi}{2}\right).
\label{fid2}
\eeq
Substituting the no-signaling probability distribution of Eq.~\eqref{step probability} in place of $p(\phi)$ in 
Eq.~\eqref{fid2} one obtains 
\eq
\bar f=1-\frac{N+1}{2\pi}\int_{\frac{-\pi}{N+1}}^{\frac{\pi}{N+1}} \sin^2\left(\frac{\phi}{2}\right).
\label{fid3}
\eeq
In the limit of large $N$ the limits of integration in Eq.~\eqref{fid3} become narrower and we can use the small angle 
approximation to write $\sin(\phi/2)\approx \phi/2$.  Substituting the latter into Eq.~\eqref{fid3} and evaluating the 
integral one obtains the average fidelity $\bar f\approx1-\frac{\pi^2}{12N^2}$.  The maximum average fidelity achievable by a quantum mechanical strategy is $\bar{f}\approx 1- \frac{\pi^2}{ 4 N^2}$~\cite{Berry:00} achieved by the input state
\eq
\label{Berry Wisemann state}
\ket{\Phi^N_{a.f.}} \propto \sum_n \sin\left(\frac{n+1}{(N+2)\pi}\right) \ket{n}.
\eeq
 This fidelity is strictly 
smaller than the bound achieved by no-signaling.  Nevertheless, the no-signaling bound gives rise to the right scaling  
with respect to $N$.

\subsection{Correspondance between asymptotic cloning and phase estimation}
\label{Assymptotic Cloning}

Every estimation strategy can be used in a measure and prepare cloning protocol (henceforth referred to as m\&p), where 
Bob first estimates the $N$-copy input state and, based on his estimate, prepares an $M$-qubit state. There are 
two free choices in every m\&p protocol: (a) what is the optimal estimation strategy, i.e., which figure of merit to chose, and,
(b) which output state to prepare, i.e., do we prepare M-copies of $\ket{\psi(\theta)}$ or some suitable $M$-partite entangled 
state. Similarly, there are two figures used in the literature thus far, to quantify the quality of a quantum cloning machine: (a) 
the \emph{global fidelity}~\cite{Ch:13} defined by the overlap of the M-qubit output of the cloning machine, $\rho^M$, with the 
the ideal $M$-copy state $\ket{\psi}^{\otimes M}$ (this is the figure of merit that we considered in the previous section) and 
(b) the \emph{per copy fidelity}~\cite{Gisin:98,Ghosh:99,Bruss:00,Buzek:96, Gisin:97,Bruss:98,Werner:98} which is the 
average of the overlap of the reduced single qubit output state 
$\rho^M|_n= \tr_{\textrm{all} \setminus n} \rho^M$ with a perfect single-qubit state $\ket{\psi}$.

In this subsection we discuss how the optimal  m\&p strategies compare with optimal cloning when the number of copies 
$M$ goes to infinity. In particular, we will show that the optimal m\&p strategy based on the \emph{alignment fidelity} is 
equivalent to an asymptotic quantum cloning machine which optimizes the \emph{per copy fidelity} of the clones, whereas 
the optimal m\&p strategy based on the maximum likelihood is equivalent to an asymptotic quantum cloning machine which 
optimizes the \emph{global fidelity}.   Whereas the equivalence between asymptotic cloning, quantified by the {\it per copy} 
fidelity, and a corresponing m\&p protocol was known for both deterministic~\cite{Bae:06} as well as probabilistic 
cloning~\cite{Gendra:14} the same question concerning asymptotic cloning,
quantified by the \emph{global} fidelity, has been an open problem~\cite{Yang:13}.


\subsubsection{Per copy cloning fidelity and alignement fidelity}

We begin by discussing the equivalence between a m\&p strategy based on the alignment fidelity and an asymptotically 
optimal cloning machine that maximizes the \emph{per copy} fidelity.  As the equivalence between a deterministic m\&p 
strategy and the optimal {\it per copy} cloning fidelity is known~\cite{Bae:06} we focus on the more general case of a 
probabilistic m\&p strategy.  The latter directly translates into the alignment fidelity of the estimation strategy as the optimal 
output state simply consists of preparing copies of $\ket{\psi(\theta')}$, where $\theta'$ is Bob's estimate of $\theta$. 
Consequently, the best m\&p average per copy cloning fidelity, given by 
$\int \,p(\theta) p(\theta'|\theta) \tr \prjct{\psi(\theta')} \prjct{\psi(\theta)} d\theta\,d\theta'$, equals the alignment fidelity 
$ \int \,p(\theta)p(\theta'|\theta) \cos^2(\frac{\theta-\theta'}{2}) d\theta\,d\theta'$. Hence, the optimal phase alignment protocol~\cite{Berry:00}, achieved for the input state 
Eq.\eqref{Berry Wisemann state}, directly translates into the optimal probabilistic m\&p cloning with per copy fidelity
$\bar f=1-\frac{\pi^2}{4 N^2}$.  Again this probabilistic strategy provides a drastic improvement over the optimal deterministic cloning strategy, where the average per copy fidelity is  $f= 1-\frac{1}{N}$ in the large $N$ limit~\cite{D'Ariano:03}.

\subsubsection{Global cloning fidelity and maximal likelihood}
Let us now turn to the global fidelity. The naive m\&p strategy consists in preparing $M$-copies 
$\ket{\psi(\theta')}^{\otimes M}$ all pointing in the estimated direction $\theta'$. In this case the output state is 
\begin{equation}\label{m&pstate}
\rho^M =\frac{1}{2\pi}\int\,d\theta\,d\theta' p(\theta'|\theta) \prjct{\psi(\theta)}^{\otimes M} d\theta.
\end{equation} 
The cloning fidelity is now given by 
\begin{equation}\label{m&p fidelity1}
F^{N\to M}_{m\&p}=\left|\frac{1}{2\pi}\int\,d\theta\,d\theta'\,p(\theta'|\theta)|\braket{\psi(\theta')}{\psi(\theta)}|^{2M}\right|
\end{equation}
Now in the limit $M\to\infty$ the overlap 
$|\braket{\psi(\theta')}{\psi(\theta)}|^{2M} \to \frac{2\sqrt{\pi}}{\sqrt{M}}\delta(\theta-\theta')$ where the constant of 
proportionality is obtained by integrating over the entire range of either $\theta$ or $\theta'$.  Inserting this expression back 
into Eq.~\eqref{m&p fidelity1} yields the global fidelity for this m\&p protocol of 
\begin{equation}\label{m&p fidelity2}
F^{N\to M}_{m\&p}=\sqrt{\frac{1}{\pi M}}p(\theta|\theta).
\end{equation}

The global fidelity of this m\&p protocol is directly proportional to the maximum likelihood for phase estimation. However, we 
note that when one substitutes the optimal maximal likelihood $p(\theta|\theta)=N+1$, achieved by the input state in 
Eq.~\eqref{ml state}, Eq.~\eqref{m&p fidelity2} is smaller than the global fidelity (Eq.~\eqref{asymptoticfidelity}), which
we proved to be the optimal fidelity achievable by the no-signaling condition, by a factor 
of $\sqrt{2}$.  A similar discrepancy was already noted in~\cite{Yang:13} for deterministic cloning, where the authors also 
showed for two simple cases (with small $N$) how to build m\&p strategies that attain the optimal asymptotic global fidelity. 
This was done by allowing Bob to output more general states. 

In the following we derive the optimal probabilistic m\&p strategy that attains the asymptomatic global fidelity of the 
probabilistic phase-covariant cloner (Eq.~\eqref{asymptoticfidelity}) for an arbitrary number of input copies $N$. We consider 
a m\&p protocol based on maximum likelihood estimation (we shall discuss its optimality in the end of the section) but allow 
Bob to output a general state $U_\theta^{\otimes M}\ket{\Psi^M}$.  

Without loss of generality let us assume that $\theta=0$.  The strategy discussed above would let Bob output the state  
\begin{equation}\label{m&p state2}
\rho_0^{m\&p} =\int p(\theta|0) U_\theta^M \prjct{\Psi^M} U_\theta^{M \dag} d\theta, 
\end{equation}
where the probability distribution $p(\theta| 0)=\tr \prjct{\Phi^N_{m.l.}}  E(\theta)=\frac{1}{(N+1)2\pi} \sum_{n,\bar n=0}^N 
e^{i \theta(n-\bar n)}$.  Here the optimal POVM elements are known to be covariant~\cite{Ch:04} and are given by 
$E(\theta)=\sum_{n,m=0}^N e^{i\theta(n-m)}\coh{n}{m}$. The corresponding  m\&p cloning fidelity is 
$F_{m\&p}^{N\to M} = \tr \rho_0^{m\&p}  \prjct{\psi(0)}^{\otimes M}$. Using the cyclic property of the trace to shift the action of 
the unitaries $U_\theta^M$ onto  $\prjct{\psi(0)}^{\otimes M}$ and carrying out the integration the global cloning fidelity reads 
\eq\label{m&p fidelity3}
F_{m\&p}^{N\to M} = \tr \prjct{\Psi^M} \mathcal{O}_M^N,
\eeq
where 
\eq
\mathcal{O}_M^N = \frac{1}{2^M}\sum_{m,\bar m=0}^M \sqrt{\binom{M}{m}\binom{M}{\bar m}}\coh{m}{\bar m} \Delta_N(
m-\bar m)
\eeq
with the coherence decay term $\Delta_N(m-\bar m)$ given by 
\begin{align}\nonumber\label{Delta terms}
&\Delta_N(m-\bar m) = \int e^{i \theta (m-\bar m)} \bra{\Phi_{m.l}^N}E(\theta)\ket{\Phi_{m.l}^N } \frac{d\theta}{2\pi} =\\ &\int \sum_{n,\bar n=0}^N e^{i\theta(n+m - \bar n -\bar m)} \frac{d\theta}{2\pi(N+1)}
=\max\{1-\frac{|m-\bar m|}{N+1},0\}.
\end{align}

Having established the form of the m\&p fidelity (Eq.~\eqref{m&p fidelity3}) we can now proceed to optimize this expression 
and obtain the corresponding optimal state $\ket{\Psi^M}$.  First, we note that the optimal state should have maximal support 
over those values of $m$ that lie in the interval $(\frac{M}{2}\pm\sqrt{\frac{M}{4}})$ since as for this range of values the 
binomial coefficients in $O_M^N$ are large. Second, $\ket{\Psi^M}$ should be roughly constant in the range $[m,m+N+1]$ 
such that that all the coherence terms, $\coh{m}{\bar m}$, add up constructively, 
i.e., $\sum_{\bar m-m} \Delta_N(m-\bar m)=N+1$.  In the limit $M\to \infty$ both of these requirements can be satisfied 
simultaneously.  In particular, choosing  $|\braket{m}{\Psi^M}|^2=\frac{1}{\sqrt{2\pi} \sigma}e^{\frac{-(m-M/2)}{2\sigma^2}}$ 
leads to a m\&p fidelity of  
\eq\label{m&p asymptotic}
F_{m\&p}^{N\to M} = \frac{(N+1) \sqrt{2}}{\sqrt{\pi M}}(1+ O(\frac{N+1}{\sigma})^2 + O(\frac{\sigma}{\sqrt{M}})^2).
\eeq
We note that Eq.~\eqref{m&p asymptotic} corresponds to the asymptotic expansion of the optimal cloning fidelity of
Eq.~\eqref{asymptoticfidelity}.  Choosing $\sigma=M^{\frac{1}{2}-\epsilon}$, for $0\leq\epsilon<\frac{1}{2}$ yields the optimal
m\&p fidelity that is equivalent to the asymptotically optimal cloning machine whose performance is quantified by the global 
fidelity.  

Note that the entire argument above is applicable even if one considers a different estimation strategy, i.e., a different figure 
of merit.  Indeed, the only thing that changes if one changes the estimation strategy (going to a general input  state $\ket{\Phi^N}$) is the coherence decay 
$\Delta_N(m-\bar m)$ in 
Eq.~\eqref{Delta terms}.  However its contribution of the coherence terms to the cloning fidelity $\sum_{m} \Delta_N(m) = p_{\ket{\Phi^N}}(0|0)$ is given by the maximal likelihood, which establishes a correspondence between the asymptotic global fidelity of the m\&p cloner and the 
maximal likelihood of the estimation (remark also that the optimal input state Eq.~\eqref{ml state} and Eq.~\eqref{deterministic} match 
for $M\to \infty$). Of course the same correspondence holds for deterministic cloning, for which the maximal likelyhood for $N
$-copies state $\ket{\psi(\theta)}^{\otimes N}$ is simply obtained as  $\bra{\psi(\theta)}^{\otimes N} E(\theta) \ket{\psi(\theta)}
^{\otimes N}= \frac{1}{2^N}(\sum_{j=0}^N \sqrt{\binom{N}{j}})^2$ and the optimal global fidelity is known to be $\frac{1}{2^{N
+M}} \binom{M}{M/2}(\sum_{j=0}^N \sqrt{\binom{N}{j}})^2 $ \cite{D'Ariano:03}.

To summarize (see table), for phase-covariant cloning the m\&p strategy based on maximal likelihood 
estimation (Sec.~\ref{sec: maximal likelihood}) is optimal with respect to the \emph{global} cloning fidelity whereas 
the m\&p strategy based on the alignment fidelity of estimation (Sec.~\ref{sec: alignment fidelity}) is optimal with respect to
the per copy fidelity of cloning. Both m\&p strategies attain the optimal cloning for any fixed $N$ and $M \to \infty$, and this is 
true both for optimal deterministic as well as probabilistic cloning. We believe that same correspondence should hold for 
universal cloning, however this is  beyond the scope of this paper. The correspondence between the different m\&p strategies
and optimal cloning machines is summarized in the following table

\begin{table}[h]
\begin{tabular}{|c c c|}
\hline
Estimation scenario for  &\vline& Optimal asymptotic cloning  \\
M\&P cloning  &\vline&  (probabilistic  or deterministic \\ \hline
Maximal likelihood & $\to$&  Global fidelity  \\
Alignment fidelity & $\to$&  Per copy  fidelity\\ \hline
\end{tabular}
\end{table}

\subsection{Metrology with general prior}
\label{sec:diffusion}
Let us now consider a more general metrological scenario where Bob has some prior knowledge, $p(\theta)$, of the 
parameter $\theta$. Following~\cite{Demkowicz:11} we consider the prior, 
$p(\theta; t) =\frac{1}{2\pi}\big(1 +2 \sum_{n=1}^\infty\cos(n\theta)e^{-n^2 t}\big)$, that arises from a diffusive 
evolution of $p(\theta)=\delta(\theta)$. The mean fidelity (Eq.~\eqref{meanfid}) now reads
\eq
\label{mean fidelity prior}
\bar f_t= 1-\int\mathrm{d}\theta' \int p(\theta'|\theta) p(\theta; t) \sin^2\left(\frac{\theta-\theta'}{2}\right) \mathrm{d}\theta
\eeq
An efficient algorithm optimizing $\bar f_t$ for moderate $N$ was derived in~\cite{Demkowicz:11}.  However, the 
optimization becomes intractable, even numerically, when $N$ increases. Indeed, optimizing $\bar f_t$ for large $N$ is 
in general a hard task.  Nevertheless the no-signaling constraint allows us to derive an upper bound for $\bar f_t$ for 
large enough $N$ as we now show.  

Our goal is to minimize the integrand of Eq.~\eqref{mean fidelity prior} under the no-signaling constraint of 
Eq.~\eqref{ns alignment}. For a fixed value of $t$ the product 
\eq
g(\theta;\theta', t)=p(\theta; t) \sin^2\left(\frac{\theta-\theta'}{2}\right)
\eeq
in Eq.~\eqref{mean fidelity prior} obtains its minimum value when $\theta-\theta'=0$. In addition $g(\theta;\theta',t)$ is 
monotonically increasing  so long as the derivative of $g(\theta;\theta',t)$ around $\theta'=\theta$  is greater than zero.  
This is true so long as 
\eq
\tan^2\left(\frac{\theta-\theta'}{2}\right) < \Big(\frac{m}{M}\Big)^2 := \tan^2(\frac{\Delta}{2}),
\eeq
where $m\equiv\min_\theta p(\theta; t)$ and $M =\max_\theta | \partial_\theta p(\theta; t)|$. Outside the interval 
$[\theta' -\Delta, \theta'+\Delta] $ the 
function $g(\theta;\theta',t)$ is larger than $m \sin^2(\Delta/2)$. Therefore, $g(\theta;\theta',t)$ attains its global 
minimum in the finite interval satisfying the condition
\eq
\sin^2\left(\frac{\theta-\theta'}{2}\right) <m \sin^2\left(\frac{\Delta}{2}\right)=\frac{m^3}{M^2+m^2}.
\eeq

Now consider the narrowest probability distribution compatible with no-signaling given by  
$\frac{N+1 }{2 \pi}p(\theta')$, where $p(\theta')$ is the probability distribution given in Eq.~\eqref{ns alignment}, for 
$|\hat \theta_\ell -\theta|<\frac{\pi}{N+1}$  and zero elsewhere. For large enough $N$ this probability distribution is 
contained entirely in the interval $[\theta' -\Delta, \theta'+\Delta] $ where $g(\theta;\theta', t)$ attains its minimum and 
therefore minimizes the integrant of Eq.~\eqref{mean fidelity prior}. Plugging this probability into 
Eq.~(\ref{mean fidelity prior}) and using the condition 
$\int \mathrm{d}\theta' p(\theta') =1$ leads to 
\eq
\bar f_t \approx 1 - \frac{\pi^2}{12 N^2} \vartheta_4(0, e^{-t}),
\eeq 
where $\vartheta_4(0, e^{-t}) = 1 + 2\sum_{n=1}^\infty (-1)^n e^{-n^2 t}= m$ is the Jacobi theta function ranging from $0$, 
when $p(\theta;0)=\delta(\theta)$, to $1$, when $p(\theta;\infty)= 1/2\pi$.  Again we discover that the ultimate bound in precision scales inversely proportional to $N^2$.
\begin{figure}[ht!]
\includegraphics[width=6cm]{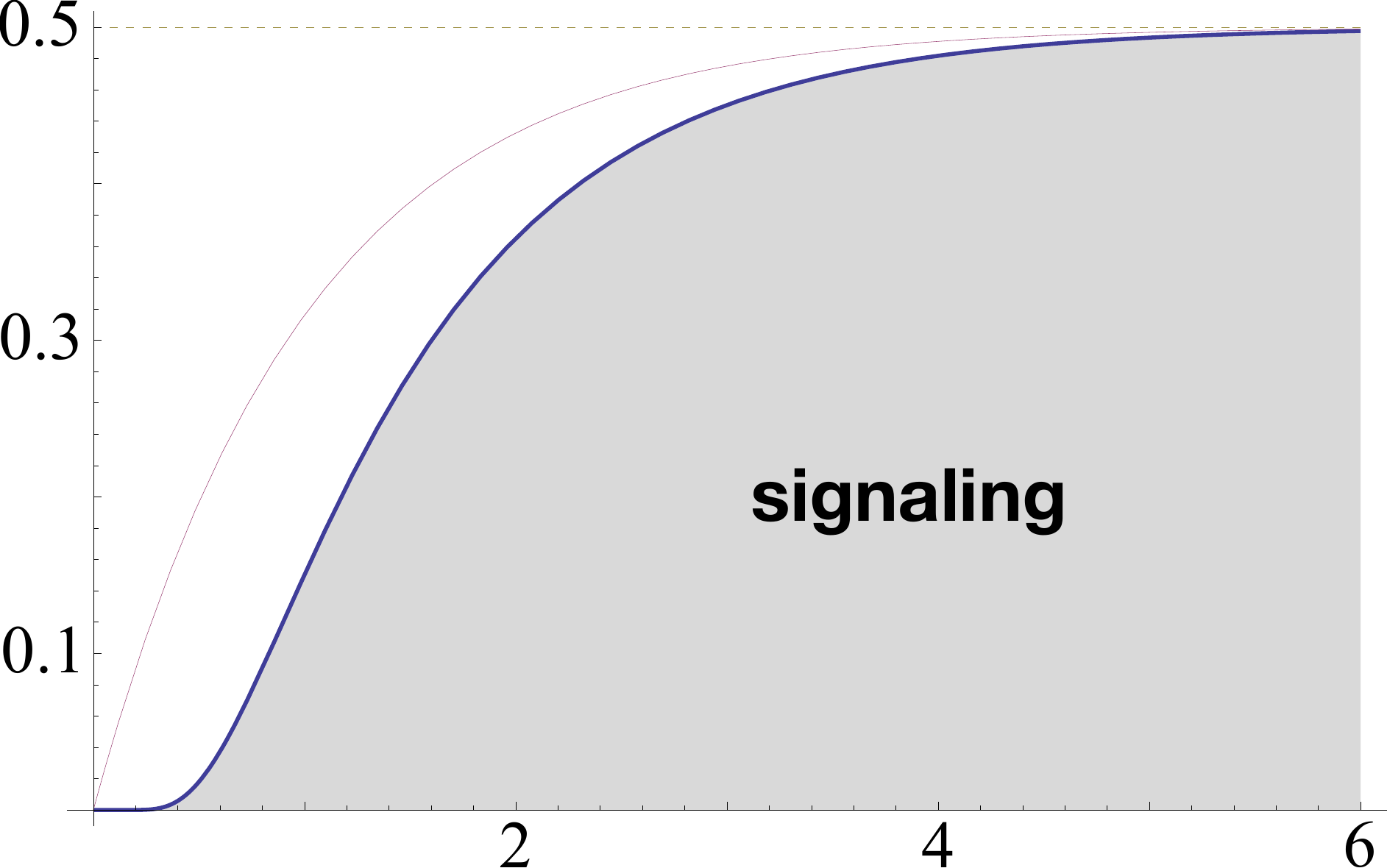}
\caption{The lower bound on the asymptotically achievable error $\frac{1}{2}\vartheta_4(0,e^{-t})= \frac{6 N^2}{\pi^2}(1-\bar f_t)$ (bottom curve) and the error of the prior $\int  p(\theta; t) \sin^2(\theta/2) d\theta$ (top curve) as functions of $t$.}
\label{fig4}
\end{figure}

\section{Discussion}
\label{sec: Discussion}

\subsection{Tightness of bounds}
We have shown that the no-signaling condition can set upper bounds on several important quantum information 
tasks, such as cloning, unitary replication, and metrology.  In the case of PCC and unitary replication we have 
shown that the no-signaling bound coincides with the optimal quantum mechanical strategy implying that quantum 
mechanical strategies for PCC cloning and unitary replication are at the edge of no-signaling.  However, for the case 
of metrology, and in particular for the average fidelity of estimation, we see that there is a gap between the 
no-signaling bound and the optimal quantum strategy.  Could this gap be an indication of the existence of a 
supra-quantum strategy, compatible with no-signaling, that outperforms the best quantum mechanical strategy?  The 
answer is no, as we now explain.

In deriving the no-signaling constraint of Eq.~\eqref{ns general} we only considered one particular way for Alice and 
Bob to attempt for faster-than-light communication; using a suitably entangled state $\ket{\Psi}_{AB}$. This, in turn 
led to the sharp probability distribution of Eq.~\eqref{step probability}.  However, one can construct a communication 
scenario where the probability distribution of Eq.~\eqref{step probability} can lead to signaling as we now show.

\begin{figure}[ht!]
\includegraphics[width=2.5cm]{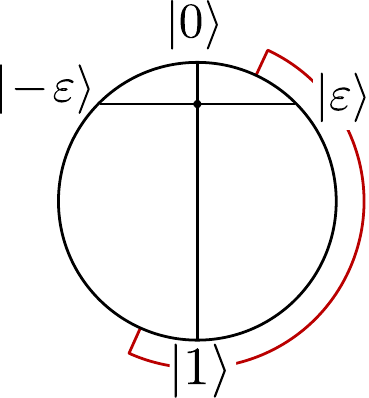}
\caption{The two ensemble decompositions of $\rho_\varepsilon$ for a qubit, leading to faster-than-light communication for the probability distribution Eq.~\eqref{step probability} (represented by the red semi-circle).}
\label{fig3}
\end{figure}

Let us first consider the qubit case ($N=1$). Let Alice and Bob share the entangled state 
$\ket{\Psi}_{AB} = \cos(\varepsilon)\ket{00} +\sin(\varepsilon)\ket{11}$. Alice can chose to measure her system in 
either the computational basis $\{\ket{0},\ket{1}\}$ or the x-basis  $\{\ket{\pm}=\frac{\ket{0}\pm \ket{1}}{\sqrt{2} }\}$ 
steering Bob's state into the ensembles $\mathcal{E}^{(1)}=\{ \cos^2(\varepsilon) \prjct{0},  \sin^2(\varepsilon)
\prjct{1}\}$ and $\mathcal{E}^{(2)}=\{\frac{1}{2} \prjct{\varepsilon}, \frac{1}{2} \prjct{-\varepsilon}\}$ respectively, where 
$\ket{\pm \varepsilon} =\cos(\varepsilon)\ket{0} \pm \sin(\varepsilon)\ket{1}$, as shown in Fig.~\ref{fig3}. 

This construction obviously holds if all the states are rotated by the same angle. In particular, we can always set this 
angle such that the probability distribution in Eq.~\eqref{step probability} yields $p(\theta'| \ket{0})= p(\theta'| \ket{-
\varepsilon})=0$ and  $p(\theta'| \ket{1})= p(\theta'| \ket{\varepsilon})=\frac{1}{\pi}$. In this case the two ensembles 
give a different probability to observe the outcome $\theta'$, $p(\theta'| \mathcal{E}^{(1)}) =\frac{\sin^2(\varepsilon)}
{\pi}$ and $p(\theta'| \mathcal{E}^{(2)}) =\frac{1}{2 \pi}$. Hence, Bob can distinguish the two ensembles with non-zero 
probability and infer Alice's choice of measurement instantaneously.

Let us now consider the general case. Any probability distribution $p(\theta)$ defines a continuous ensemble $\mathcal{E}^{(p)} =\{ p(\theta) \prjct{\Phi_\theta^N}\}$, where $\ket{\Phi_\theta^N}=\sum_{n=0}^N \psi_n e^{i \theta n}\ket{n}$  are the $N$-qubit states of Eq.~\eqref{state}. Without loss of generality we consider $p(\theta)$ such that 
\eq\label{FT}
p(\theta) =\frac{1}{2\pi}(1+ 2 \sum_{k=1}^\infty p_k \cos(k \theta)).
\eeq
The density matrix for the ensemble $\mathcal{E}^{(p)}$ is given by $\rho = \sum_{n,m =0}^N \psi_n \psi_m^* \coh{n}{m} p_{|n-m|}$, in such a way that it only depends on the first $N$ coefficients, $p_k$, of the Fourier series in Eq.~\eqref{FT}. For any two distributions $p_1(\theta)$ and $p_2(\theta)$ that are identical in the first $N$ components of the Fourier series the ensembles $\mathcal{E}^{(p_1)}$ and $\mathcal{E}^{(p_2)}$ give rise to the same density matrix $\rho$, and therefore can not be distinguished by Bob.

In particular, the ensemble given by the probability distributions $p_1(\theta)= \frac{1}{2\pi}$ and $p_2(\theta) = \frac{1}{2\pi}(1 + \cos(M \theta))$, for $M>N$, correspond to the same density matrix. However, with the outcome probability distribution of Eq.~\eqref{step probability} Bob can distinguish the two probability distributions with non-zero probability as $p(\theta'| \mathcal{E}^{(p_2)}) - p(\theta'| \mathcal{E}^{(p_1)}) = \frac{N+1}{2\pi }\int_{-\frac{\pi}{N+1}}^{\frac{\pi}{N+1}}(p_2(\theta)-p_1(\theta))d\theta = \mathrm{sinc}(\frac{M\pi}{N+1})\neq 0$. Therefore, the probability distribution of Eq.~\eqref{step probability} leads to signaling when Alice can chose to prepare  $\mathcal{E}^{(p_1)}$ or  $\mathcal{E}^{(p_2)}$. 

More generally the above argument implies that any outcome probability $p(\theta'| \theta)$ compatible with no-signaling has to satisfy $\int p(\theta'|\theta)\cos(M\theta)d\theta =0$ for $M>N$, i.e., the Fourier components $p_k$ of $p(\theta'| \theta)$ are necessarily zero for $k>N$. Therefore, for finite $N$, probability distributions with sharp edges such as the one in  Eq.~\eqref{step probability} are ruled out.

A tighter no-signaling bound can be obtained if 
we optimize over all possible no-signaling scenarios, i.e., over all possible bi-partite entangled states 
$\ket{\Psi}_{AB}$. In fact any ensemble $\{p_k, \rho_k\}$ corresponding to a density matrix, $\rho_B = \sum_k p_k \rho_k$, at Bob's side can be remotely prepared by Alice, if they initially share a suitable entangled state $\ket{\Psi}_{AB}$ (that only depends on $\rho_B$) and Alice does an appropriate measurement~\cite{Gisin:89, Hughston:93}.

\subsection{Quantum Mechanics at the edge of no-signaling}
The above argument shows that the probability distribution of Eq.~\eqref{step probability} is only valid for one 
possible no-signaling scenario, and that in order to obtain a tighter bound we  should consider all possible states shared 
between Alice and Bob and all possible measurements at Alice's side that steer Bob's partial state into different 
ensembles of pure states that correspond to the same density matrix.  Would such an optimization close the gap between our no-signaling 
bound and the optimal quantum strategy?  

Following \cite{Simon:01}, we now show that such an optimization is not even necessary, as the only 
processing compatible with no-signaling is given by the Born rule, i.e., the probability of some measurement outcome $\ell$ for the input state $\rho$ is given by $P_\ell = \tr \rho E_\ell$ for some positive operator $E_\ell$.

Indeed, any ensemble leading to the same density matrix for Bob can be remotely prepared by Alice~\cite{Gisin:89, Hughston:93}.  This together with the no-signaling condition implies the linearity of Bob's processing, $\mathcal{P}$.  The latter states that for any 
two ensembles $\{p_k, \, \rho_k \}$ and $\{q_k, \,  \sigma_k \}$, corresponding to the same $\rho_B$, the ensembles after the processing 
$\{p_k, \, \mathcal{P}(\rho_k) \}$ and $\{q_k, \, \mathcal{P}(\sigma_k) \}$ can not be distinguished with a non-zero probability. Adding the assumption that probabilities are attributed to quantum states via the Born rule (as it is done in \cite{Simon:01}) the condition above implies equality on the density matrices corresponding to the processed ensembles:
\eq\label{linearity1}
\sum_k p_k \mathcal{P}(\rho_k) = \sum_k q_k \mathcal{P}(\sigma_k) \equiv \mathcal{P}(\rho_B).
\eeq

This  shows that any  dynamical evolution of quantum states that respects no-signaling is necessarily described by a 
completely positive 
(CP) map~\cite{Simon:01}. The results of~\cite{Simon:01} are concerned with situations where the outputs of Bob's 
processing are quantum mechanical states. In this case \cite{Simon:01} implies that the optimal quantum 
mechanical strategies are at the edge of no-signaling.  

However, in the case of quantum metrology the outputs are probability distributions. We remark than in~\cite{Simon:01} the 
validity of the Born rule 
was assumed and used to derive the possibility for remote state preparation of any ensemble and to get the linearity 
constraint of Eq.~\eqref{linearity1} from the indistinguishably of processed ensembles.  In this case supra-quantum metrology 
is ruled out.  However, if we make no assumptions on how probabilities are assigned to measurement outcomes of quantum 
states but only take remote state preparation as an experimental fact, the no-signaling constraint implies the Born rule 
already. In this case no-signaling again implies the indistinguishably of two ensembles 
$\{p_k, \, \rho_k \}$ and $\{q_k, \,  \sigma_k \}$ corresponding to the same density matrix $\rho_B$ which in turn implies the 
linearity of the probability assignment rule. The probability $P_\ell$ to observe some outcome $\ell$ has to satisfy 
\eq \label{linearity2}
\sum_k p_k P_\ell (\rho_k) = \sum_k q_k P_\ell (\sigma_k) \equiv P_\ell( \rho_B),
\eeq
i.e., outcome probabilities \emph{only depend on the density matrix} but not on a particular ensemble. Note that one can 
easily construct probability assignment rules for pure states that do not satisfy Eq.~\eqref{linearity2} (see the example from 
the previous section) so it is not something one has to impose a priori. However, as we just saw assuming no-signaling 
together with the practical possibility for steering enforces linearity.

It is well known that the only probability assignment rule compatible with linearity is the Born 
rule---$P_\ell(\rho_B)=\tr \rho_B E_\ell$ for some positive operator $E_\ell$~\cite{Holevo:82}. For systems of  dimension 
$d>2$ this can also be seen as a consequence of Gleason's theorem (it suffices to consider all ensembles of pure states 
forming an orthonormal basis). Moreover, similar result for the equivalence of CP dynamics and the Born rule being enforced 
by linearity are known to hold in a more general context of probabilistic theories with purification~\cite{Ch:10}. In summary, 
we have shown that also in the case of quantum metrology the optimal quantum mechanical strategies are at the edge of 
no-signaling. 

\subsection{Probabilistic vs deterministic bounds} 
Notice that all the no-signaling bounds derived here are concerned with probabilistic strategies.  This is most 
transparent in our derivation of a no-signaling bound for unitary replication. In some cases, the optimal deterministic 
strategy coincides with the optimal probabilistic one as is the case with replication of unitaries. In other cases the optimal 
probabilistic strategy can be made deterministic if one drops some restrictions on the input alphabet, as is the case for the 
cloning of states when one allows for entangled input states $U_\theta^{\otimes N} \ket{\Phi^N}$ instead of separable input 
states $\ket{\psi(\theta)}^{\otimes N}$. How does one decide if a deterministic bound still holds when one allows for 
probabilistic tasks? If this is not the case can one achieve the probabilistic performance deterministically by allowing for more 
general input states? 
 
It is known that, in general, any probabilistic strategy can be decomposed into a filter, $F$, acting on the input state and 
mapping it to an output state in the same Hilbert space, followed by a deterministic transformation~\cite{Ch:13}. 
Moreover, one is usually interested in probabilistic strategies where all states from the input alphabet $\{ \ket{\Phi_i}\}$ have 
the same probability of success $p_s = \tr F \prjct{\Phi_i}  F^\dag$. In this case what the probabilistic advantage has to offer 
is the possibility  to modify the alphabet to any other alphabet reachable by a filter 
$\{\ket{\Phi_i^F}=\frac{1}{\sqrt{p_s}} F \ket{\Phi^F_i} \}$~\footnote{If the input alphabet is generated by the action of a unitary 
on a fixed seed state $\ket{\Psi_i} = U_i \ket{\Psi}$, then the filter commutes with the unitary and modifies the seed 
state to $\ket{\Psi^F} =\frac{1}{\sqrt{p_s}} F \ket{\Psi}$}. So the question about the strength of the deterministic bound is 
actually whether the input alphabet $\{ \ket{\Phi_i}\}$ is the best among alphabets  $\{\ket{\Phi_i^F}\}$ for the particular task. 

If this is not the case, then the probabilistic strategy can always be made deterministic by starting with the optimal alphabet. 
As we saw in Sec.~\ref{sec:PCC} for the case of PCC the $N$-copy input states are not optimal leaving room for probabilistic 
improvement, whereas in the case of universal cloning no probabilistic advantage exists as the symmetry group of the input 
alphabet, $SU(2)$, forces the filter to be the identity.  In fact, there is a substantial improvement in cloning fidelity if one 
allows entangled $N$-qubit states as inputs into the cloning machine, but such states are not reachable by any 
filter~\cite{Ch:13}.  The entangled states that yield such a substantial improvement are exactly those states that maximize the 
average fidelity of alignment for a Cartesian frame of reference~\cite{Ch:04a}.
\newline
\section{Conclusion}

In this paper we derived no-signaling bounds for various quantum information processing tasks.  These include  
phase covariant cloning of states and unitary operations, as well as quantum metrology. In the latter case we showed 
the validity of the Heisenberg limit purely from the no-signaling condition.  In general, following~\cite{Simon:01}, we have 
shown that the optimal probabilistic quantum mechanical strategy is at the edge of no-signaling also for the case of 
metrology.  Furthermore, we have found that for some tasks, such as PCC of states and unitaries, the optimal probabilistic 
and deterministic strategies coincide.  These results show a direct connection between the no-signalling principle and the 
ultimate limits on quantum cloning and metrology. This connection provides a new insight into the {\em physical} origin of 
these limits, in contrast to the previously known limits based on optimization, using e.g. semidefinte programs.  

On the one hand it is clear that a bound for probabilistic strategies is also a bound for deterministic ones. However, it 
might be possible to derive tighter no-signaling bounds for deterministic strategies. It is an interesting open question 
how to incorporate the requirement that the protocol be deterministic in a no-signaling scenario. 

On the other hand, there are several tasks for which the optimal quantum strategy is not known.  In such cases the 
techniques and methods we provide here can be particularly useful in deriving limitations to these tasks based on 
no-signaling. We have demonstrated one such example for Bayesian metrology for arbitrary prior, however the methods we 
introduce are applicable in a broader context. This provides an alternative approach to study the possibilities and limitations 
of quantum information processing.

\section{Acknowledgements}

We thank Giulio Chiribella, Nicolas Gisin, and Christoph Simon for valuable remarks and comments. This work was 
supported by the Austrian Science Fund (FWF): P24273-N16 and the Swiss National Science Foundation grant 
P2GEP2\_151964.
\bibliographystyle{apsrev4-1}
\bibliography{nosignaling}

\end{document}